\documentclass[12pt,letterpaper,doublespace]{article}
\usepackage[T1]{fontenc}
\usepackage[utf8]{inputenc}
\usepackage{prettyref}
\usepackage{amstext}
\usepackage[numbers]{natbib}

\makeatletter
\newcommand{\lyxaddress}[1]{
\par {\raggedright #1
\vspace{1.4em}
\noindent\par}
}

\@ifundefined{date}{}{\date{}}
\usepackage{prettyref}
\newrefformat{eq}{eq.\nolinebreak(\ref{#1})}
\newrefformat{sec}{Section\nolinebreak\,\nolinebreak\ref{#1}}
\newrefformat{cha}{Chapter\nolinebreak\,\nolinebreak\ref{#1}}
\newrefformat{tab}{Table\nolinebreak\,\nolinebreak\ref{#1}}
\newrefformat{fig}{Figure\nolinebreak\,\nolinebreak\ref{#1}}
\newrefformat{lem}{Lemma\nolinebreak\,\nolinebreak\ref{#1}}
\newrefformat{thm}{Theorem\nolinebreak\,\nolinebreak\ref{#1}}
\newrefformat{axm}{Axiom\nolinebreak\,\nolinebreak\ref{#1}}
\newrefformat{app}{Appendix\nolinebreak\,\nolinebreak\ref{#1}}
\newrefformat{exe}{Exercise\nolinebreak\,\nolinebreak\ref{#1}}
\newrefformat{cor}{Corollary\nolinebreak\,\nolinebreak\ref{#1}}
\newrefformat{pro}{Property\nolinebreak\,\nolinebreak\ref{#1}}
\newrefformat{def}{Definition\nolinebreak\,\nolinebreak\ref{#1}}
\newrefformat{subsec}{Section\nolinebreak\,\nolinebreak\ref{#1}}
\newrefformat{fn}{Footnote\nolinebreak\,\nolinebreak\ref{#1}}
\usepackage{amsmath}
\usepackage{bm}
\numberwithin{equation}{section}
\newcommand{\br}[1]{\mathbf{#1}} 
\newcommand{\bs}[1]{\bm{\mathsf{#1}}} 
\newcommand{\bb}[1]{\mathbb{#1}} 
\newcommand{\ca}[1]{\mathcal{#1}} 
\newcommand{\bg}[1]{\bm{#1}} 
\newcommand{\dy}[1]{\bb{#1}} 
\newcommand{\fv}[1]{\bs{#1}} 

\usepackage{pifont}
\usepackage{txfonts}
\makeatother

\begin{document}

\title{Magnetic Charge and Dyality Invariance}

\author{Oliver Davis Johns}
\maketitle

\lyxaddress{San Francisco State University, Department of Physics and Astronomy,
Thornton Hall, 1600 Holloway Avenue, San Francisco, CA 94132 USA,
<ojohns@metacosmos.org>}
\begin{abstract}
This paper is a critical study of non-standard Maxwellian electrodynamics.
It explores two important topics: the inclusion of both magnetic and
electric charge to produce what it calls \emph{Extended Electrodynamics,}
and the existence of a symmetry called \emph{Dyality Invariance} that
exchanges electric and magnetic quantities.\emph{ }

First, the paper summarizes \emph{Extended Electrodynamics,} including
potentials, gauge transformations, and a new proof of the extended
electrodynamic Poynting theorem. 

A formal Lagrangian derivation of the extended Maxwell equations is
also given, but its value in fundamental studies is questioned.

The paper then defines \emph{Dyality Invariance} (form invariance
under the so-called \emph{Dyality Transformation} that exchanges electric
and magnetic quantities) and shows it to be a valid symmetry if and
only if electrodynamics is given the \emph{extended} form.

The paper suggests that the complete Maxwellian electrodynamics is
extended electrodynamics with its dyality invariance. But dyality
can be interpreted either \emph{actively} or \emph{passively.} Since
magnetic charge has not been observed experimentally, the active interpretation
is ruled out. But a passive interpretation can be used to avoid writing
magnetic source and potential terms explicitly.

The paper also refutes the idea that dyality invariance would permit
a magnetic charge to be transformed away even if one existed. If nonzero
magnetic charge exists, then experimental evidence for its existence
cannot be hidden by a dyality transformation.\pagebreak{}
\end{abstract}

\section{Preface \label{sec:Preface}}

This paper explores two important topics: the inclusion of both magnetic
and electric charge to produce what is called \emph{Extended Electrodynamics,}
and the existence of a symmetry called \emph{Dyality Invariance} that
exchanges electric and magnetic quantities.\footnote{The neologism \textquotedbl{}dyality\textquotedbl{} is suggested by
\citep{han-dyality} to prevent confusion with \textquotedbl{}duality\textquotedbl{},
the relation, for example, in \prettyref{eq:du1}. }

Sections \ref{sec:notation} through \ref{sec:Gauge} summarize \emph{Extended
Electrodynamic}s, including potentials and gauge transformations.
A new proof of the extended electrodynamic Poynting theorem is given
in \prettyref{sec:ElecMag} and Appendix A in \prettyref{sec:AppendixA}. 

\prettyref{sec:Lagrange} gives a formal Lagrangian derivation of
extended electrodynamics, but suggests that Lagrangian methods are
of limited value.

In \prettyref{sec:Trans},  \emph{Dyality Invariance} (form invariance
under an exchange of electric and magnetic quantities called a \emph{Dyality
Transformation}) is defined and shown to be a symmetry \emph{only}
of extended electrodynamics. Standard electrodynamics with only electric
charge is not invariant under dyality transformations. 

Sections \ref{sec:Trans} and \ref{sec:DualStructure} demonstrate
the dyality invariance of extended electrodynamics, including the
extended Maxwell equations themselves, the energy-momentum tensor,
and expressions involving the four-vector potentials.

\prettyref{sec:DualityImply} introduces the distinction between \emph{active}
and \emph{passive} interpretations of dyality transformations. Active
interpretation of dyality invariance would imply the experimental
existence of magnetic charge, and is therefore currently ruled out.
However, since magnetic charge has not yet been found experimentally,
a passive interpretation of dyality invariance allows us to consider
the standard electrodynamics to be extended electrodynamics with the
magnetic sources and potentials transformed passively out of all equations.

\prettyref{sec:DualityImply} also offers a refutation of the idea
that dyality invariance would permit an \emph{experimentally existing}
magnetic charge to be transformed away. If nonzero magnetic charge
exists, experimental evidence for its existence cannot be hidden by
any dyality transformation.

An afterword in \prettyref{sec:Afterword} suggests directions for
future research. The paper has two Appendices.

\section{Notation and Definitions\label{sec:notation}}

This section introduces some definitions to be used in the paper. 

An event $x^{\mu}$ is denoted by
\begin{align}
x^{0} & =ct\quad\quad x^{\mu}=(x^{0},x^{1},x^{2},x^{3})\nonumber \\
\partial_{\mu} & =\left(\dfrac{\partial}{\partial x^{0}},\dfrac{\partial}{\partial x^{1}},\dfrac{\partial}{\partial x^{2}},\dfrac{\partial}{\partial x^{3}}\right)\quad\:\square^{2}=\partial_{\mu}\partial^{\mu}\label{eq:use1}
\end{align}
We denote four-vectors as $\fv K=K^{0}\fv e_{0}+\br K$ where $\fv e_{0}$
is the time unit vector and the three-vector part is understood to
be $\br K=K^{1}\fv e_{1}+K^{2}\fv e_{2}+K^{3}\fv e_{3}$. In the Einstein
summation convention, Greek indices range from 0 to 3, Roman indices
from 1 to 3. The Minkowski metric tensor used to raise or lower indices
is $\eta_{\mu\nu}=\eta^{\mu\nu}=\text{diag}(-1,+1,+1,+1)$. Three-vectors
are written with bold type $\br K$, and their magnitudes as $K$.
The paper uses Heaviside-Lorentz electromagnetic units, and considers
only electrodynamics in a vacuum except for explicit source charge
densities.

The completely antisymmetric, Minkowski-space, Levi-Civita tensor
$\varepsilon^{\alpha\beta\mu\nu}$ obeys the identities
\begin{equation}
\varepsilon^{0123}=+1\quad\quad\varepsilon_{0123}=-1\label{eq:use2}
\end{equation}
\begin{equation}
\varepsilon^{\alpha\beta\mu\nu}\varepsilon_{\alpha\beta\gamma\delta}=-2\left(\delta_{\gamma}^{\mu}\delta_{\delta}^{\nu}-\delta_{\delta}^{\mu}\delta_{\gamma}^{\nu}\right)\label{eq:use3}
\end{equation}
\begin{equation}
\varepsilon^{\alpha\beta\mu\nu}\varepsilon_{\alpha\theta\gamma\delta}=-\left(\delta_{\theta}^{\beta}\delta_{\gamma}^{\mu}\delta_{\delta}^{\nu}+\delta_{\gamma}^{\beta}\delta_{\delta}^{\mu}\delta_{\theta}^{\nu}+\delta_{\delta}^{\beta}\delta_{\theta}^{\mu}\delta_{\gamma}^{\nu}\right)+\left(\delta_{\theta}^{\beta}\delta_{\delta}^{\mu}\delta_{\gamma}^{\nu}+\delta_{\delta}^{\beta}\delta_{\gamma}^{\mu}\delta_{\theta}^{\nu}+\delta_{\gamma}^{\beta}\delta_{\theta}^{\mu}\delta_{\delta}^{\nu}\right)\label{eq:use4}
\end{equation}
where $\delta_{\beta}^{\alpha}=\eta^{\alpha\mu}\eta_{\beta\mu}$ is
the Kroeneker-delta tensor, which is +1 when $\alpha=\beta$ and zero
otherwise.

If $X^{\alpha\beta}$ is an antisymmetric second rank tensor, then
its \emph{dual }$\tilde{X}^{\alpha\beta}$ is another antisymmetric
second rank tensor, defined as 
\begin{equation}
\tilde{X}^{\alpha\beta}=\dfrac{1}{2}\varepsilon^{\alpha\beta\mu\nu}X_{\mu\nu}\label{eq:use5}
\end{equation}
It follows from \prettyref{eq:use3} that
\begin{equation}
\tilde{\tilde{X}}^{\alpha\beta}=\dfrac{1}{2}\varepsilon^{\alpha\beta\mu\nu}\tilde{X}_{\mu\nu}=\frac{1}{4}\varepsilon^{\alpha\beta\mu\nu}\varepsilon_{\mu\nu\theta\delta}X^{\theta\delta}=-\frac{1}{2}\left(\delta_{\theta}^{\alpha}\delta_{\delta}^{\beta}-\delta_{\delta}^{\alpha}\delta_{\theta}^{\beta}\right)X^{\theta\delta}=-X^{\alpha\beta}\label{eq:use5a}
\end{equation}
Other useful identities follow from eqs.( \ref{eq:use4} and \ref{eq:use5}).
If $X^{\alpha\beta}$ and $Y^{\alpha\beta}$ are antisymmetric tensors,
then
\begin{equation}
\tilde{X}^{\alpha\mu}\tilde{Y}_{\alpha\nu}=Y^{\alpha\mu}X_{\alpha\nu}-\frac{1}{2}\delta_{\nu}^{\mu}X^{\alpha\beta}Y_{\alpha\beta}\label{eq:use6}
\end{equation}
\begin{equation}
\tilde{X}^{\alpha\beta}\tilde{Y}_{\alpha\beta}=-X^{\alpha\beta}Y_{\alpha\beta}\label{eq:use6a}
\end{equation}

The Maxwell field four-tensor $F^{\alpha\beta}$ in terms of the electric
field $\br E$ and the magnetic field $\br B$ is
\begin{equation}
F^{\mu\nu}=\left(\begin{array}{cccc}
0 & E_{x} & E_{y} & E_{z}\\
-E_{x} & 0 & B_{z} & -B_{y}\\
-E_{y} & -B_{z} & 0 & B_{x}\\
-E_{z} & B_{y} & -B_{x} & 0
\end{array}\right)_{\mu\nu}\label{eq:intro2}
\end{equation}
The equations of electrodynamics can often be put in more concise
and symmetric form by also defining a dual field tensor $G^{\alpha\beta}$
as 
\begin{equation}
G^{\alpha\beta}=\tilde{F}^{\alpha\beta}=\dfrac{1}{2}\varepsilon^{\alpha\beta\mu\nu}F_{\!\!\mu\nu}\label{eq:du1}
\end{equation}
 The inverse relation can be found from \prettyref{eq:use5a} 
\begin{equation}
F^{\alpha\beta}=-\tilde{\tilde{F}}^{\alpha\beta}=-\tilde{G}^{\alpha\beta}=-\dfrac{1}{2}\varepsilon^{\alpha\beta\mu\nu}G_{\mu\nu}\label{eq:du3}
\end{equation}
In terms of the electric field $\br E$ and the magnetic field $\br B$,
the dual field tensor is
\begin{equation}
G^{\alpha\beta}=\left(\begin{array}{cccc}
0 & B_{x} & B_{y} & B_{z}\\
-B_{x} & 0 & -E_{z} & E_{y}\\
-B_{y} & E_{z} & 0 & -E_{x}\\
-B_{z} & -E_{y} & E_{x} & 0
\end{array}\right)_{\alpha\beta}\label{eq:du4}
\end{equation}

The standard Maxwell equations without magnetic charge then have the
manifestly covariant form\footnote{See eq.(12.126) of \citep{Griffiths}.}
\begin{equation}
\partial_{\alpha}F^{\alpha\beta}=-\,\dfrac{1}{c}\,J^{\beta}\quad\text{and}\quad\partial_{\alpha}G^{\alpha\beta}=0\label{eq:du5}
\end{equation}
where $J^{\alpha}$ is the electric charge density four-vector with
electric charge density $\rho$ and flux density $\br J$. The three-vector
form of the standard Maxwell equations is
\begin{equation}
\bg\nabla\cdot\br E=\rho\quad\quad\bg\nabla\times\br B=+\dfrac{1}{c}\dfrac{\partial\br E}{\partial t}+\dfrac{1}{c}\br J\label{eq:du6}
\end{equation}
\begin{equation}
\bg\nabla\cdot\br B=0\quad\quad-\bg\nabla\times\br E=\dfrac{1}{c}\dfrac{\partial\br B}{\partial t}\label{eq:du7}
\end{equation}
Inserting \prettyref{eq:du1} into the second of \prettyref{eq:du5}
gives
\begin{equation}
\varepsilon^{\beta\alpha\mu\nu}\partial_{\alpha}F_{\!\!\mu\nu}=0\label{eq:du8}
\end{equation}
for all $\beta$ values. This equation is equivalent to 
\begin{equation}
\partial_{\alpha}F_{\!\!\mu\nu}+\partial_{\mu}F_{\!\!\nu\alpha}+\partial_{\nu}F_{\!\!\alpha\mu}=0\label{eq:du9}
\end{equation}
which is often quoted\footnote{See eq.(11.143) of \citep{Jackson}.}
as the covariant form of the so-called homogeneous Maxwell equations,
\prettyref{eq:du7}. However, the second of \prettyref{eq:du5} itself
seems the preferable form since it shows clearly the absence of a
magnetic charge density parallel to the electric charge density $J^{\beta}$.

\section{Extended Electrodynamics with Electric and Magnetic Charge}

\label{sec:ElecMag}In this section we present a synopsis of an \emph{Extended
Electrodynamics }with both electric and magnetic charges, and prove
an extended electrodynamic version of the Poynting theorem.

Assuming that magnetic charge, if any, must be added to the Maxwell
equations in a way that preserves Lorentz covariance, a conservative
and plausible generalization of standard electrodynamics is simply
to add a magnetic charge density four-vector as source of the dual
field tensor in \prettyref{eq:du5}.

The covariant Maxwell equations are then:
\begin{equation}
\partial_{\alpha}F^{\alpha\beta}=-\,\dfrac{1}{c}\,J^{\beta}\quad\quad\partial_{\alpha}G^{\alpha\beta}=-\dfrac{1}{c}L^{\beta}\label{eq:em1}
\end{equation}
where
\begin{equation}
\fv J=c\rho\fv e_{0}+\br J\quad\quad\fv L=c\lambda\fv e_{0}+\br L\label{eq:em2}
\end{equation}
are, respectively, the electric charge density four-vector $\fv J$
with electric charge density $\rho$ and flux density $\br J$ and
the magnetic charge four-vector $\fv L$ with magnetic charge density
$\lambda$ and flux density $\br L$. 

In three-vector form, the extended Maxwell equations in \prettyref{eq:em1}
are\footnote{Eqs.(\ref{eq:em3}, \ref{eq:em4}) agree with eq.(6.150) of \citep{Jackson}.}
\begin{equation}
\bg\nabla\cdot\br E=\rho\:\:\quad\bg\nabla\times\br B=\dfrac{1}{c}\dfrac{\partial\br E}{\partial t}+\dfrac{1}{c}\br J\label{eq:em3}
\end{equation}
\begin{equation}
\bg\nabla\cdot\br B=\lambda\:\:\quad-\bg\nabla\times\br E=\dfrac{1}{c}\dfrac{\partial\br B}{\partial t}+\dfrac{1}{c}\br L\label{eq:em4}
\end{equation}

It follows from \prettyref{eq:em1} that, due to the anti-symmetry
of $F^{\alpha\beta}$ and $G^{\alpha\beta}$,
\begin{equation}
\partial_{\beta}J^{\beta}=-c\partial_{\alpha}\partial_{\beta}F^{\alpha\beta}=0\quad\quad\partial_{\beta}L^{\beta}=-c\partial_{\alpha}\partial_{\beta}G^{\alpha\beta}=0\label{eq:em5}
\end{equation}
In three-vector form these are the conservation rules for electric
and magnetic charge
\begin{equation}
\dfrac{\partial\rho}{\partial t}+\bg\nabla\cdot\br J=0\quad\quad\dfrac{\partial\lambda}{\partial t}+\bg\nabla\cdot\br L=0\label{eq:em6}
\end{equation}

Another plausible generalization defines the extended Lorentz force
density four-vector $f^{\alpha}$ by adding a comparable magnetic
term $f_{\text{mg}}^{\alpha}$ to the standard electric term $f_{\text{el}}^{\alpha}$
so that
\begin{equation}
f^{\alpha}=\left(f_{\text{el}}^{\alpha}+f_{\text{mg}}^{\alpha}\right)\quad\quad\text{where}\quad\quad f_{\text{el}}^{\alpha}=\dfrac{1}{c}F_{\:\:\gamma}^{\alpha}J^{\gamma}\quad\quad f_{\text{mg}}^{\alpha}=\dfrac{1}{c}G_{\:\:\gamma}^{\alpha}L^{\gamma}\label{eq:em16}
\end{equation}
In three-vector form,
\begin{equation}
f_{\text{el}}^{0}=\dfrac{1}{c}\left(\br E\cdot\br J\right)\quad\quad\quad f_{\text{mg}}^{0}=\dfrac{1}{c}\left(\br B\cdot\br L\right)\label{eq:em17}
\end{equation}
\begin{equation}
\br f_{\text{el}}=\left\{ \rho\br E+\dfrac{1}{c}\left(\br J\times\br B\right)\right\} \quad\quad\quad\br f_{\text{mg}}=\left\{ \lambda\br B-\dfrac{1}{c}\left(\br L\times\br E\right)\right\} \label{eq:em18}
\end{equation}
We can also use \prettyref{eq:em1} to write the force density entirely
in terms of field tensors
\begin{align}
f_{\text{el}}^{\alpha} & =\dfrac{1}{c}F_{\:\:\gamma}^{\alpha}J^{\gamma}=-F_{\:\:\gamma}^{\alpha}\left(\partial_{\mu}F^{\mu\gamma}\right)\label{eq:em19}\\
f_{\text{mg}}^{\alpha} & =\dfrac{1}{c}G_{\:\:\gamma}^{\alpha}L^{\gamma}=-G_{\:\:\gamma}^{\alpha}\left(\partial_{\mu}G^{\mu\gamma}\right)\label{eq:em20}
\end{align}

Standard electrodynamics defines\footnote{For example, see eq.(12.113) of \citep{Jackson}.}
the symmetric energy momentum tensor as 
\begin{equation}
T^{\alpha\beta}=F^{\alpha\mu}F^{\beta\nu}\,\eta_{\mu\nu}\,-\,\dfrac{1}{4}\,\eta^{\alpha\beta}F_{\!\!\mu\nu}F^{\mu\nu}\label{eq:intro1}
\end{equation}
This same definition also proves correct for extended electrodynamics. 

Expansion of \prettyref{eq:intro1} using matrix multiplication yields
\begin{equation}
T^{\alpha\beta}=T^{\beta\alpha}=\left(\begin{array}{cccc}
{\cal E} & c\,P_{x} & c\,P_{y} & c\,P_{z}\\
c\,P_{x} & M_{11} & M_{12} & M_{13}\\
c\,P_{y} & M_{21} & M_{22} & M_{23}\\
c\,P_{z} & M_{31} & M_{32} & M_{33}
\end{array}\right)_{\alpha\beta}\label{eq:intro3}
\end{equation}
where,
\begin{equation}
{\cal E}=\frac{1}{2}\left(E^{2}+B^{2}\right)\quad\quad c\br P=\br E\times\br B\quad\quad M_{ij}=-\left(E_{i}E_{j}+B_{i}B_{j}\right)\,+\delta_{ij}\ca E\label{eq:intro4}
\end{equation}
The second term on the right in \prettyref{eq:intro1} has the effect
of making $T^{\alpha\beta}$ traceless. With the invariant trace defined
as
\begin{equation}
\text{Tr}\:T=T^{\alpha\beta}\eta_{\alpha\beta}\label{eq:intro5}
\end{equation}
it follows from \prettyref{eq:intro1} that 
\begin{equation}
\text{Tr}\:T=\eta_{\alpha\beta}\eta_{\mu\nu}F^{\alpha\mu}F^{\beta\nu}-F_{\!\!\mu\nu}F^{\mu\nu}=0\label{eq:intro6}
\end{equation}

Using the identity \prettyref{eq:use6} with the substitutions  $X^{\alpha\beta}=Y^{\alpha\beta}=F^{\alpha\beta}$,
the standard definition in \prettyref{eq:intro1} can also be written
in an equivalent form
\begin{equation}
T^{\alpha\beta}=G{}^{\alpha\mu}G{}^{\beta\nu}\,\eta_{\mu\nu}\,-\,\dfrac{1}{4}\,\eta^{\alpha\beta}G{}_{\mu\nu}G{}^{\mu\nu}\label{eq:du14}
\end{equation}
Since the same substitutions show that $G_{\mu\nu}G{}^{\mu\nu}=-F_{\mu\nu}F{}^{\mu\nu}$,
eqs.(\ref{eq:intro1} and \ref{eq:du14}) can be added to give a third
equivalent form
\begin{equation}
T^{\alpha\beta}=\dfrac{1}{2}\left\{ F^{\alpha\mu}F^{\beta\nu}\,\eta_{\mu\nu}+G^{\alpha\mu}G^{\beta\nu}\,\eta_{\mu\nu}\right\} \label{eq:du13}
\end{equation}
Of these three equivalent forms, the third, \prettyref{eq:du13},
is the simplest and most useful. It should be quoted in the textbooks
rather than \prettyref{eq:intro1}. 

As will be seen in \prettyref{sec:Trans}, \prettyref{eq:du13} makes
evident the invariance of $T^{\alpha\beta}$ under the dyality transformation.
The otherwise accidental requirement that $T^{\alpha\beta}$ must
be made traceless before it will reproduce the Poynting theorem can
be understood as the requirement that electrodynamics must be invariant
under the dyality transformation, and hence must be extended electrodynamics.

Using the equivalent definition of $T^{\alpha\beta}$ from \prettyref{eq:du13},
together with eqs.(\ref{eq:em19}, \ref{eq:em20}), Appendix A in
\prettyref{sec:AppendixA} proves that
\begin{equation}
\partial_{\alpha}T^{\alpha\beta}=-\left(f_{\text{el}}^{\beta}+f_{\text{mg}}^{\beta}\right)=-f^{\beta}\label{eq:em21}
\end{equation}
which demonstrates both the correctness of the choice of $T^{\alpha\beta}$
in the equivalent eqs.(\ref{eq:intro1}, \ref{eq:du14}, and \ref{eq:du13}),
and also the correctness of the force hypothesis in \prettyref{eq:em16}. 

For $\beta=0$, \prettyref{eq:em21} expands to the Poynting theorem
\begin{equation}
\left(\dfrac{\partial{\cal E}}{\partial t}+\bg\nabla\cdot\br S\right)=-\left(\br E\cdot\br J\right)-\left(\br B\cdot\br L\right)\label{eq:em22}
\end{equation}
where $\br S=c\br E\times\br B$. 

For $\beta=i$, where $i=1,2,3$, \prettyref{eq:em21} expands to
\begin{equation}
\dfrac{\partial P_{i}}{\partial t}+\left(\bg\nabla\cdot\dy M\right)_{i}=-\left\{ \rho\br E+\dfrac{1}{c}\left(\br J\times\br B\right)\right\} _{i}-\left\{ \lambda\br B-\dfrac{1}{c}\left(\br L\times\br E\right)\right\} _{i}\label{eq:em23}
\end{equation}
 where $\br P$ and $\dy M$ are defined in \prettyref{eq:intro4}.
Note that the sign of the three-dimensional dyadic $\dy M$ is defined
here so that, with $d\br a$ the outward pointing elements of surface
${\cal S}$, the integral $\oint_{{\cal S}}\left(\sum_{j=1}^{3}M_{ij}da_{j}\right)$
is the net \emph{outgoing }flow of the $i^{\text{th}}$ component
of momentum. 

\section{Extended Maxwell Equations Derived from Two Vector Potentials}

\label{sec:potentials}The extended Maxwell equations can be derived
from two four-vector potentials, $\fv A$ and $\fv N$. With the definition\footnote{Eqs.(\ref{eq:pot1}, \ref{eq:pot2a}, \ref{eq:ga7}, \ref{eq:ga9})
were first obtained by Shanmugadhasan\citep{Shanmugadhasan}. See
also \citep{cabibbo-ferrari}. For sign convention, compare eq.(11.136)
of \citep{Jackson}.}
\begin{equation}
F^{\alpha\beta}=\left(\dfrac{\partial A^{\beta}}{\partial x^{\alpha}}-\dfrac{\partial A^{\alpha}}{\partial x^{\beta}}\right)-\varepsilon^{\alpha\beta\mu\nu}\left(\dfrac{\partial N^{\nu}}{\partial x^{\mu}}-\dfrac{\partial N^{\mu}}{\partial x^{\nu}}\right)\label{eq:pot1}
\end{equation}
it follows that the dual field tensor is
\begin{equation}
G^{\alpha\beta}=\dfrac{1}{2}\varepsilon^{\alpha\beta\mu\nu}F_{\!\!\mu\nu}=\frac{1}{2}\varepsilon^{\alpha\beta\mu\nu}\left(\partial_{\mu}A_{\nu}-\partial_{\nu}A_{\mu}\right)-\frac{1}{2}\varepsilon^{\alpha\beta\mu\nu}\varepsilon_{\mu\nu\gamma\delta}\partial^{\gamma}N^{\delta}\label{eq:pot2}
\end{equation}
Using the identity \prettyref{eq:use3}, this is
\begin{equation}
G^{\alpha\beta}=\left(\dfrac{\partial N^{\beta}}{\partial x^{\alpha}}-\dfrac{\partial N^{\alpha}}{\partial x^{\beta}}\right)+\varepsilon^{\alpha\beta\mu\nu}\left(\dfrac{\partial A^{\nu}}{\partial x^{\mu}}-\dfrac{\partial A^{\mu}}{\partial x^{\nu}}\right)\label{eq:pot2a}
\end{equation}

With the definitions of the antisymmetric and gauge invariant tensors
$a_{\alpha\beta}$ and $n_{\alpha\beta}$,
\begin{flalign}
a_{\alpha\beta} & =\dfrac{\partial A_{\alpha}}{\partial x^{\beta}}-\dfrac{\partial A_{\beta}}{\partial x^{\alpha}}\nonumber \\
n_{\alpha\beta} & =\dfrac{\partial N_{\alpha}}{\partial x^{\beta}}-\dfrac{\partial N_{\beta}}{\partial x^{\alpha}}\label{eq:lag12}
\end{flalign}
and the definition of duals in \prettyref{eq:use5}, the eqs.(\ref{eq:pot1}
and \ref{eq:pot2a}) may also be written as
\begin{flalign}
F_{\alpha\beta} & =-a_{\alpha\beta}+\tilde{n}_{\alpha\beta}\label{eq:lag13}\\
G_{\alpha\beta} & =-n_{\alpha\beta}-\tilde{a}_{\alpha\beta}\nonumber 
\end{flalign}

The potential four-vectors may be written with the notations $A^{0}=\phi$
and $N^{0}=\theta$
\begin{equation}
\fv A=\phi\fv e_{0}+\br A\quad\quad\quad\fv N=\theta\fv e_{0}+\br N\label{eq:pot7}
\end{equation}
Then the electric and magnetic fields can be written in terms of these
potentials.
\begin{equation}
-E_{i}=F^{i0}=\partial^{i}A^{0}-\partial^{0}A^{i}-\varepsilon^{i0kl}\partial_{k}N_{l}\label{eq:pot8}
\end{equation}
and thus
\begin{equation}
\br E=-\bg\nabla\phi-\frac{1}{c}\dfrac{\partial\br A}{\partial t}-\bg\nabla\times\br N\label{eq:pot9}
\end{equation}
Also
\begin{equation}
-B_{i}=G^{i0}=\partial^{i}N^{0}-\partial^{0}N^{i}+\varepsilon^{i0kl}\partial_{k}A_{l}\label{eq:pot10}
\end{equation}
and thus
\begin{equation}
\br B=-\bg\nabla\theta-\frac{1}{c}\dfrac{\partial\br N}{\partial t}+\bg\nabla\times\br A\label{eq:pot11}
\end{equation}

\section{Gauge Transformation of Potentials\label{sec:Gauge}}

A gauge transformation replaces the four-vector potentials $A^{\alpha}$
and $N^{\alpha}$ introduced in \prettyref{sec:potentials} by the
modified potentials 
\begin{equation}
A^{*\alpha}=A^{\alpha}+\partial^{\alpha}\Lambda\quad\text{and}\quad N{}^{*\alpha}=N^{\alpha}+\partial^{\alpha}\Gamma\label{eq:ga1}
\end{equation}
where $\Lambda$ and $\Gamma$ are field functions. Then
\begin{equation}
F^{*\alpha\beta}=\left(\partial^{\alpha}A{}^{*\beta}-\partial^{\beta}A{}^{*\alpha}\right)-\varepsilon^{\alpha\beta\mu\nu}\partial_{\mu}N{}_{\nu}^{*}=F^{\alpha\beta}+\left(\partial^{\alpha}\partial^{\beta}-\partial^{\beta}\partial^{\alpha}\right)\Lambda-\frac{1}{2}\varepsilon^{\alpha\beta\mu\nu}\left(\partial_{\mu}\partial_{\nu}-\partial_{\nu}\partial_{\mu}\right)\Gamma=F^{\alpha\beta}\label{eq:ga2}
\end{equation}
 and similarly $G^{*\alpha\beta}=G^{\alpha\beta}$. The field tensors
$F^{\alpha\beta}$and $G^{\alpha\beta}$, and thus the electric and
magnetic fields $\br E$ and $\br B$, are unchanged by gauge transformation
of the potentials.

We assume the Lorenz conditions $\partial_{\alpha}A^{\alpha}=0$ and
$\partial_{\alpha}N^{\alpha}=0$ for the four-vector potentials. The
gauge-transformed potentials $A^{*\alpha}$ and $N^{*\alpha}$ are
also four-vectors satisfying the Lorenz conditions $\partial_{\alpha}A^{*\alpha}=0$
and $\partial_{\alpha}N^{*\alpha}=0$ if and only if the field functions
$\Lambda$ and $\Gamma$ are Lorentz scalar fields satisfying 
\begin{equation}
\square^{2}\Lambda=0\quad\text{and}\quad\square^{2}\Gamma=0\label{eq:ga5}
\end{equation}
which we also assume here.

Then the first of \prettyref{eq:em1} can be written as
\begin{equation}
-\frac{1}{c}J^{\beta}=\partial_{\alpha}F^{\alpha\beta}=\partial_{\alpha}\partial^{\alpha}A^{\beta}-\partial^{\beta}\left(\partial_{\alpha}A^{\alpha}\right)-\varepsilon^{\alpha\beta\gamma\delta}\partial_{\alpha}\partial_{\gamma}N_{\delta}\label{eq:ga6}
\end{equation}
The last term vanishes due to antisymmetry and hence, assuming the
Lorenz gauge condition, $\partial_{\alpha}A^{\alpha}=0$, 
\begin{equation}
\square^{2}A^{\beta}=-\frac{1}{c}J^{\beta}\label{eq:ga7}
\end{equation}
Similarly, assuming the Lorenz gauge condition $\partial_{\alpha}N^{\alpha}=0$
and the second of \prettyref{eq:em1},
\begin{equation}
-\frac{1}{c}L^{\beta}=\partial_{\alpha}G^{\alpha\beta}=\partial_{\alpha}\partial^{\alpha}N^{\beta}-\partial^{\beta}\left(\partial_{\alpha}N^{\alpha}\right)-\varepsilon^{\alpha\beta\gamma\delta}\partial_{\alpha}\partial_{\gamma}A_{\delta}\label{eq:ga8}
\end{equation}
\begin{equation}
\square^{2}N^{\beta}=-\frac{1}{c}L^{\beta}\label{eq:ga9}
\end{equation}

\section{Formal Lagrangian Derivation of \protect \\
Extended Electrodynamics}

\label{sec:Lagrange}This section shows that the Maxwell equations
of extended electrodynamics can be derived from a Lagrangian field
function. 

Some approaches to a Lagrangian derivation of extended electrodynamics
have been merely the starting point of an attempt to derive the Dirac
monopole by merging Lagrangian electrodynamics with Dirac spinor sources.\footnote{For example, \citep{Shanmugadhasan,Fryberger,cabibbo-ferrari}.}

These earlier attempts at a Lagrangian for extended electrodynamics
are constricted by their attempt to write the Lagrangian somehow as
a sum of separate \textquotedbl{}electric\textquotedbl{} and \textquotedbl{}magnetic\textquotedbl{}
Lagrangians ${\cal L}={\cal L}_{\text{el}}+{\cal L}_{\text{mg}}$.
Others\footnote{See \citep{Rohrlich-1966-Mono}.} introduce \textquotedbl{}electric\textquotedbl{}
fields $F^{\alpha\beta}$ that act only on electric charges and \textquotedbl{}magnetic\textquotedbl{}
fields $F'^{\alpha\beta}$ that act only on magnetic charges. 

The simplification here is, first, that the Lagrangian field ${\cal L}$
need not be separated into a symmetric sum of electric and magnetic
Lagrangians. 

The second simplification is that only one kind of electromagnetic
field $F^{\alpha\beta}$ is needed. It can be \emph{expressed }also
in the form of a dual tensor $G^{\alpha\beta}=\tilde{F}^{\alpha\beta}$,
but the $\br E$ and $\br B$ fields used in the two tensors are the
same, just re-arranged. One must avoid conflating dyality and duality. 

Although a formal Lagrangian derivation of the extended Maxwell equations
proves possible, we must remember that application of Lagrangian methods
to electromagnetism has a strong element of enlightened guesswork.
We are loosely guided by analogy, and the precise choices $d/dt\rightarrow\partial/\partial x^{\mu}$,
$u_{k}\rightarrow A_{\alpha}$, $v_{k}\rightarrow\Phi_{\alpha\mu}=\partial A_{\alpha}/\partial x^{\mu}$
suggested by mechanics must be justified by their success in practice. 

The simplified Lagrangian chosen here is\footnote{\label{fn:moulin}See \citep{Moulin}. Since \prettyref{eq:use6a}
with $X{=}Y{=}F$ implies that $G^{\alpha\beta}G_{\alpha\beta}=-F^{\alpha\beta}F_{\alpha\beta}$,
this Lagrangian can also be written in the more symmetric form ${\cal L}=-\left\{ \frac{1}{8}F^{\alpha\beta}F_{\alpha\beta}+\frac{1}{c}J^{\alpha}A_{\alpha}\right\} +\left\{ \frac{1}{8}G^{\alpha\beta}G_{\alpha\beta}+\frac{1}{c}L^{\alpha}N_{\alpha}\right\} $.}
\begin{equation}
{\cal L}=-\frac{1}{4}F^{\alpha\beta}F_{\alpha\beta}-\frac{1}{c}J^{\alpha}A_{\alpha}+\frac{1}{c}L^{\alpha}N_{\alpha}\label{eq:lag1}
\end{equation}
The role of \textquotedbl{}coordinates\textquotedbl{} is played by
the potentials $A_{\alpha}$ and $N_{\alpha}$, and the role of \textquotedbl{}generalized
velocities\textquotedbl{} by $\Phi_{\alpha\beta}$ and $\Theta_{\alpha\beta}$
where
\begin{equation}
\Phi_{\alpha\beta}=\dfrac{\partial A_{\alpha}}{\partial x^{\beta}}\quad\text{and}\quad\Theta_{\alpha\beta}=\dfrac{\partial N_{\alpha}}{\partial x^{\beta}}\label{eq:lag2}
\end{equation}
To express ${\cal L}$ in terms of these \textquotedbl{}coordinates\textquotedbl{}
and \textquotedbl{}generalized velocities\textquotedbl{}, write \prettyref{eq:pot1}
and \prettyref{eq:pot2a} as
\begin{equation}
F_{\alpha\beta}=-\left(\Phi_{\alpha\beta}-\Phi_{\beta\alpha}\right)+\varepsilon_{\alpha\beta}^{\;\;\;\;\mu\nu}\Theta_{\mu\nu}\quad\quad G_{\alpha\beta}=-\left(\Theta_{\alpha\beta}-\Theta_{\beta\alpha}\right)-\varepsilon_{\alpha\beta}^{\;\;\;\;\mu\nu}\Phi_{\mu\nu}\label{eq:lag3}
\end{equation}

In taking partial derivatives, the Lagrangian field is considered
as a function of the form\\
 ${\cal L}={\cal L}\left(A,N,\Phi,\Theta,x\right)$ where $A\equiv\left[A_{\alpha}\right]$,
$N\equiv\left[N_{\alpha}\right]$, $\Phi\equiv\left[\partial A_{\alpha}/\partial x_{\mu}\right]$,
$\Theta\equiv\left[\partial N_{\alpha}/\partial x_{\mu}\right]$,
and $x\equiv\left[x^{\mu}\right]$, and where the {[} {]} brackets
denote the entire set of enclosed components. It follows that
\begin{equation}
\dfrac{\partial}{\partial\Phi_{\alpha\beta}}\left\{ F^{\lambda\delta}F_{\lambda\delta}\right\} =-2\left\{ \dfrac{\partial}{\partial\Phi_{\alpha\beta}}\left(\Phi_{\lambda\delta}-\Phi_{\delta\lambda}\right)\right\} F^{\lambda\delta}=-4F^{\alpha\beta}\label{eq:lag4}
\end{equation}
and
\begin{equation}
\dfrac{\partial}{\partial\Theta_{\alpha\beta}}\left\{ F^{\lambda\delta}F_{\lambda\delta}\right\} =2\left\{ \dfrac{\partial}{\partial\Theta_{\alpha\beta}}\left(\varepsilon_{\lambda\delta}^{\;\;\;\;\mu\nu}\Theta_{\mu\nu}\right)\right\} F^{\lambda\delta}=2\varepsilon^{\lambda\delta\alpha\beta}F_{\lambda\delta}=4G^{\alpha\beta}\label{eq:lag5}
\end{equation}
Hence
\begin{equation}
\dfrac{\partial{\cal L}\left(A,N,\Phi,\Theta,x\right)}{\partial\Phi_{\alpha\beta}}=F^{\alpha\beta}\quad\quad\quad\dfrac{\partial{\cal L}\left(A,N,\Phi,\Theta,x\right)}{\partial\Theta_{\alpha\beta}}=-G^{\alpha\beta}\label{eq:lag7}
\end{equation}
 The Lagrange equations, again chosen by analogy with mechanics, are
\begin{equation}
\partial_{\beta}\left(\dfrac{\partial{\cal L}}{\partial\Phi_{\alpha\beta}}\right)-\dfrac{\partial{\cal L}}{\partial A_{\alpha}}=0\quad\quad\quad\partial_{\beta}\left(\dfrac{\partial{\cal L}}{\partial\Theta_{\alpha\beta}}\right)-\dfrac{\partial{\cal L}}{\partial N_{\alpha}}=0\label{eq:lag8}
\end{equation}
Using eqs.( \ref{eq:lag1} and \ref{eq:lag7}), the Lagrange equations
in \prettyref{eq:lag8} expand to
\begin{equation}
\partial_{\beta}F^{\alpha\beta}-\frac{1}{c}J^{\alpha}=0\quad\quad\quad-\partial_{\beta}G^{\alpha\beta}+\frac{1}{c}L^{\alpha}=0\label{eq:lag9}
\end{equation}
Taking account of the anti-symmetry of $F^{\alpha\beta}$ and $G^{\alpha\beta}$,
\prettyref{eq:lag9} reproduces the covariant extended Maxwell equations
in \prettyref{eq:em1}.\footnote{The Lagrangian function ${\cal L}$ in \prettyref{eq:lag1} is neither
invariant under gauge transformation nor invariant under dyalitic
transformations. This is admissible because, in both cases, the resulting
Lagrange equations \prettyref{eq:lag9} are invariant. }

Note however that the Lagrangian analogy is limited. For example,
the analogs of the generalized momenta defined in mechanics as $p_{k}=\partial L/\partial\dot{q}_{k}$,
are the fields $F^{\alpha\mathbf{\beta}}$ and $-G^{\alpha\beta}$
in \prettyref{eq:lag7}. But these are not independent; they are just
the Maxwell field tensor and its dual. There is no covariant, extended
electrodynamic analog of the Hamilton equations. 

Attempts to derive the energy-momentum tensor $T^{\alpha\beta}$ from
the Lagrangian are also unpersuasive. In Section 4.9 of \citep{Rohrlich-ClassPart},
a symmetrizing correction term is added to a standard Lagrangian derivation
of the energy-momentum tensor beginning from Noether's theorem. This
method does give the correct $T^{\alpha\beta}$ for \emph{standard}
electrodynamics. The same method is quoted in \citep{Jackson} and
\citep{goldstein}. However, the generalization of this method does
not produce the correct $T^{\alpha\beta}$ for \emph{extended} electrodynamics,
even after a symmetrizing correction is added. 

Also, a Lagrangian derivation of $T^{\alpha\beta}$ in $\mathsection$94
of \citep{Landau-Fields}, using Hamilton's principle together with
variation of a general-relativistic metric, gives the correct value,
but is formalistic and unpersuasive.

The most reliable verification of the energy-momentum tensor $T^{\alpha\beta}$
quoted in \prettyref{eq:du13} is the direct, algebraic proof of the
Poynting theorem in \prettyref{sec:ElecMag} and Appendix A in \prettyref{sec:AppendixA}.

\section{The Dyality Transformation\label{sec:Trans}}

This section defines the \emph{Dyality Transformation} and shows \emph{Dyality
Invariance} under this transformation to be a symmetry only of extended
electrodynamics.

For any solution to the source-free, standard Maxwell equations (eqs.(\ref{eq:du6},
\ref{eq:du7}) with $\rho=0$ and $\br J=0$) there is an alternate
solution with primed fields defined as
\begin{equation}
\br E'=\br B\:\:\quad\br B'=-\br E\label{eq:tx1a}
\end{equation}
With these definitions, eqs.(\ref{eq:du6}, \ref{eq:du7}) give 
\begin{equation}
-\bg\nabla\cdot\br B'=0\quad\quad\bg\nabla\times\br E'=-\dfrac{1}{c}\dfrac{\partial\br B'}{\partial t}\label{eq:tx2a}
\end{equation}
\begin{equation}
\bg\nabla\cdot\br E'=0\quad\quad\bg\nabla\times\br B'=\dfrac{1}{c}\dfrac{\partial\br E'}{\partial t}\label{eq:tx3a}
\end{equation}
which shows that the primed fields satisfy exactly the same Maxwell
equations as eqs.(\ref{eq:du6}, \ref{eq:du7}), but with primes on
all fields.

However, that symmetry is broken when $\rho\neq0$ or $\br J\neq0$.
Then, for example, $-\bg\nabla\cdot\br B'=\rho$ which is not a correct
Maxwell equation. The transformation in \prettyref{eq:tx1a} is not
a valid symmetry of standard electrodynamics when sources are present. 

Symmetry can be regained by moving to the extended electrodynamics
of \prettyref{sec:ElecMag}, and including both fields and sources
in the dyality transformation. Inclusion of $\fv A$ and $\fv N$
also guarantees dyality invariance of expressions involving these
potentials. The \emph{Dyality Transformation }is then defined as
\begin{equation}
\left(\begin{array}{c}
\br E'\\
\br B'
\end{array}\right)=\left(\begin{array}{c}
\br B\\
-\br E
\end{array}\right)\quad\quad\quad\left(\begin{array}{c}
\fv J'\\
\fv L'
\end{array}\right)=\left(\begin{array}{c}
\fv L\\
-\fv J
\end{array}\right)\quad\quad\quad\left(\begin{array}{c}
\fv A'\\
\fv N'
\end{array}\right)=\left(\begin{array}{c}
\fv N\\
-\fv A
\end{array}\right)\label{eq:tx1}
\end{equation}

When these definitions are substituted into \prettyref{eq:em3} and
\prettyref{eq:em4}, the primed quantities then satisfy the same three-vector
Maxwell equations as in \prettyref{sec:ElecMag}
\begin{equation}
-\bg\nabla\cdot\br B'=-\lambda'\quad\:\:\bg\nabla\times\br E'=-\dfrac{1}{c}\dfrac{\partial\br B'}{\partial t}-\dfrac{1}{c}\br L'\label{eq:tx2}
\end{equation}
\begin{equation}
\bg\nabla\cdot\br E'=\rho'\:\:\quad\bg\nabla\times\br B'=\dfrac{1}{c}\dfrac{\partial\br E'}{\partial t}+\dfrac{1}{c}\br J'\label{eq:tx3}
\end{equation}
 This is the dyality transformed solution of the Maxwell equations
in which electric and magnetic quantities are exchanged. 

This \emph{Dyality Invariance} is a symmetry only of \emph{Extended
Electrodynamics,} with both electric and magnetic charges. As shown
above, there is no such symmetry for standard electrodynamics except
in the special source-free case.

Denote by $F'^{\alpha\beta}$ and $G'^{\alpha\beta}$ the same matrices
as in \prettyref{eq:intro2} and \prettyref{eq:du4}, but written
with primes on the electric and magnetic field components. Thus
\begin{equation}
F'^{\alpha\beta}=\left(\begin{array}{cccc}
0 & E'_{x} & E'_{y} & E'_{z}\\
-E'_{x} & 0 & B'_{z} & -B'_{y}\\
-E'_{y} & -B'_{z} & 0 & B'_{x}\\
-E'_{z} & B'_{y} & -B'_{x} & 0
\end{array}\right)_{\alpha\beta}\label{eq:tx5}
\end{equation}
with a similar expression for $G'^{\alpha\beta}$
\begin{equation}
G'^{\alpha\beta}=\left(\begin{array}{cccc}
0 & B'_{x} & B'_{y} & B'_{z}\\
-B'_{x} & 0 & -E'_{z} & E'_{y}\\
-B'_{y} & E'_{z} & 0 & -E'_{x}\\
-B'_{z} & -E'_{y} & E'_{x} & 0
\end{array}\right)_{\alpha\beta}\label{eq:tx5a}
\end{equation}
 Then substitution of the definitions in \prettyref{eq:tx1} into
\prettyref{eq:tx5} and \prettyref{eq:tx5a} gives the dyality transformation
of $F^{\alpha\beta}$ and $G^{\alpha\beta}$ as 
\begin{equation}
\left(\begin{array}{c}
F'^{\alpha\beta}\\
G'^{\alpha\beta}
\end{array}\right)=\left(\begin{array}{c}
G^{\alpha\beta}\\
-F^{\alpha\beta}
\end{array}\right)\label{eq:tx6}
\end{equation}
Substitute these into \prettyref{eq:em1} to obtain 
\begin{equation}
-\partial_{\mu}G'^{\mu\nu}=\,\dfrac{1}{c}\,L'^{\nu}\quad\text{and}\quad\partial_{\mu}F'^{\mu\nu}=-\dfrac{1}{c}J'^{\nu}\label{eq:tx7}
\end{equation}
in which the two equations in \prettyref{eq:em1} are merely interchanged.
The primed quantities thus satisfy the same covariant Maxwell equations
as those listed above in \prettyref{eq:em1}. This is the covariant
form of the alternate solution in eqs.(\ref{eq:tx2a} and \ref{eq:tx3a}).

We now consider the effect of the dyality transformation on the energy-momentum
tensor $T^{\alpha\beta}$ as written in one of its equivalent forms
in \prettyref{eq:du13}. Let $T'^{\alpha\beta}$ denote the same matrix
as the $T^{\alpha\beta}$ in \prettyref{eq:du13} but written with
primes on the field components
\begin{equation}
T'^{\alpha\beta}=\dfrac{1}{2}\left\{ F'^{\alpha\mu}F'^{\beta\nu}\,\eta_{\mu\nu}+G'^{\alpha\mu}G'^{\beta\nu}\,\eta_{\mu\nu}\right\} \label{eq:tx8}
\end{equation}
Substitution of \prettyref{eq:tx6} into \prettyref{eq:tx8} gives
\begin{equation}
T'^{\alpha\beta}=\dfrac{1}{2}\left\{ F'^{\alpha\mu}F'{}^{\beta\nu}\,\eta_{\mu\nu}+G'{}^{\alpha\mu}G'{}^{\beta\nu}\,\eta_{\mu\nu}\right\} =\dfrac{1}{2}\left\{ G^{\alpha\mu}G{}^{\beta\nu}\,\eta_{\mu\nu}+\left(-F^{\alpha\mu}\right)\left(-F^{\beta\nu}\right)\,\eta_{\mu\nu}\right\} =T^{\alpha\beta}\label{eq:tx9}
\end{equation}
The dyality transformation \prettyref{eq:tx6} simply exchanges the
two terms in \prettyref{eq:du13}. Then $T^{\alpha\beta}$ is invariant
under the dyality transformation, in the sense that the dyality substitution
produces
\begin{equation}
T'^{\alpha\beta}=T{}^{\alpha\beta}\label{eq:tx10}
\end{equation}

This invariance of $T^{\alpha\beta}$ can also be verified by inspection
of \prettyref{eq:intro3}. Note that the dyality substitution makes
\begin{equation}
E'_{i}E'_{j}+B'_{i}B'_{j}=B{}_{i}B{}_{j}+\left(-E_{i}\right)\left(-E_{j}\right)=E_{i}E{}_{j}+B{}_{i}B{}_{j}\label{eq:tx10a}
\end{equation}
 and hence ${\cal E}'={\cal E}$ and $M'_{ij}=M_{ij}$. Also notice
that $\br E'\times\br B'=\br B\times\left(-\br E\right)=\br E\times\br B$
and hence $P'_{i}=P_{i}$. Thus $T'^{\alpha\beta}=T{}^{\alpha\beta}$.

The dyality transformation must not be confused with the Lorentz transformation
that sums over indices. The equation $T'^{\alpha\beta}=T{}^{\alpha\beta}$
holds independently for every matrix element of $T^{\alpha\beta}$.
Thus \prettyref{eq:tx10} implies that $T'^{00}=T{}^{00}$, $T'^{0i}=T{}^{0i}$,
and so forth for the other indices.

\section{The Dyality Structure of Extended Electrodynamics}

\label{sec:DualStructure}One feature of the covariant form of the
extended electrodynamics in \prettyref{sec:ElecMag} is the split
of the equations into those governed by the field tensor $F^{\alpha\beta}$
with the electric charge density four-vector $\fv J$ on the one hand,
and those governed by the dual field tensor $G^{\alpha\beta}$ with
the magnetic charge density four-vector $\fv L$ on the other.

We see this division in the Maxwell equations in \prettyref{eq:em1}
where
\begin{equation}
\partial_{\alpha}F^{\alpha\beta}=-\,\dfrac{1}{c}\,J^{\beta}\quad\text{and}\quad\partial_{\alpha}G^{\alpha\beta}=-\dfrac{1}{c}L^{\beta}\label{eq:dual1}
\end{equation}
and in the Lorentz force densities in \prettyref{eq:em16} where $f^{\alpha}=(f_{\text{el}}^{\alpha}+f_{\text{mg}}^{\alpha})$
with
\begin{equation}
f_{\text{el}}^{\alpha}=\dfrac{1}{c}F_{\:\:\gamma}^{\alpha}J^{\gamma}\quad\text{and}\quad f_{\text{mg}}^{\alpha}=\dfrac{1}{c}G_{\:\:\gamma}^{\alpha}L^{\gamma}\label{eq:dual2}
\end{equation}
This split also appears in the equations for the vector potentials
$\fv A$ and $\fv N$ in \prettyref{eq:ga7} and \prettyref{eq:ga9}
where
\begin{equation}
\square^{2}A^{\beta}=-\frac{1}{c}J^{\beta}\quad\text{and}\quad\square^{2}N^{\beta}=-\frac{1}{c}L^{\beta}\label{eq:dual3}
\end{equation}
The four-vector potential $\fv A$ has only electric sources $\fv J$,
and the four-vector potential $\fv N$ has only magnetic sources $\fv L$.

In each of the split cases, eqs.(\ref{eq:dual1}, \ref{eq:dual2},
\ref{eq:dual3}), the dyality transformation exchanges the two parts
of the split, thus demonstrating the dyality invariance of the theory.

The energy momentum tensor $T^{\alpha\beta}$ in \prettyref{eq:du13}
is a single expression written as a sum of quadratic terms in $F^{\alpha\beta}$
and $G^{\alpha\beta}$
\begin{equation}
T^{\alpha\beta}=\dfrac{1}{2}\left\{ F^{\alpha\mu}F^{\beta\nu}\,\eta_{\mu\nu}+G^{\alpha\mu}G^{\beta\nu}\,\eta_{\mu\nu}\right\} \label{eq:du13-1}
\end{equation}
 The dyality transformation exchanges these terms, thus producing
the dyality invariance of $T^{\alpha\beta}$ seen in \prettyref{eq:tx10}.

The dyality transformation also exchanges the expressions in \prettyref{eq:pot1}
and \prettyref{eq:pot2a} for the Maxwell field $F^{\alpha\beta}$
and its dual $G^{\alpha\beta}$ in terms of the vector potentials;
the \prettyref{eq:pot1} 

\begin{equation}
F^{\alpha\beta}=\left(\partial^{\alpha}A^{\beta}-\partial^{\beta}A^{\alpha}\right)-\varepsilon^{\alpha\beta\mu\nu}\partial_{\mu}N_{\nu}\label{eq:dual3a}
\end{equation}
becomes \prettyref{eq:pot2a}
\begin{equation}
G^{\alpha\beta}=\left(\partial^{\alpha}N{}^{\beta}-\partial^{\beta}N{}^{\alpha}\right)+\varepsilon^{\alpha\beta\mu\nu}\partial_{\mu}A{}_{\nu}\label{eq:dual3b}
\end{equation}
and vice-versa.

\section{Does Dyality Invariance Imply a Magnetic Monopole?\label{sec:DualityImply}}

Let us tentatively accept the hypothesis that a dyality invariant
extended electrodynamics is the correct form of electrodynamics. We
consider the implications of that hypothesis. In particular, what
does it say about the reality or non-reality of magnetic charge? 

The answer depends on the interpretation of the dyality transformation:
active or passive.

What is called an \emph{active} interpretation of the dyality transformation
would imply the experimental existence of magnetic charge, and hence
is currently ruled out. 

The analogy here is with\emph{ active }rotations in standard three-dimensional
vector algebra.\footnote{See Chapter 8 of \citep{oj}} In active rotations,
a three-vector $\br V$ is transformed into a rotated vector $\br V'$
whose components in the original coordinate system are given by $V'_{i}=\sum R_{ij}V_{j}$.
The rotated vector $\br V'$ has the same length but a different direction
and hence is physically different. 

Applying that analogy to \prettyref{eq:tx1}, in an active transformation
the primed fields represent a different physical reality. Due to dyality
invariance, they are still solutions to the Maxwell equations and
the other equations of extended electrodynamics, but they are \emph{alternate}
solutions, not the same one. 

For example, a plane, monochromatic, linearly polarized light wave
in vacuum, with angular velocity $\omega$ and wave vector $\br k=\left(\omega/c\right)\fv e_{3}$
has 
\begin{equation}
\br E=a\fv e_{1}\,\cos\phi\quad\quad\br B=a\fv e_{2}\,\cos\phi\label{eq:im1a}
\end{equation}
where $\phi=\left(kz-\omega t\right)$ and $z=x^{3}$. Apply \prettyref{eq:tx1}
to obtain 
\begin{equation}
\br E'=a\fv e_{2}\,\cos\phi\quad\quad\br B'=-a\fv e_{1}\,\cos\phi\label{eq:im1b}
\end{equation}
which is a plane wave linearly polarized in the $\fv e_{2}$ direction.
The primed plane wave in \prettyref{eq:im1b} is also a solution to
Maxwell equations, but it is an \emph{alternate} solution, physically
distinguishable from the original solution in \prettyref{eq:im1a}. 

For a more complex example, suppose that we begin with the extended
Maxwell equations, but with $\fv L=0$ since no magnetic charge has
been found experimentally
\begin{equation}
\bg\nabla\cdot\br E=\rho\:\:\quad\bg\nabla\times\br B=\dfrac{1}{c}\dfrac{\partial\br E}{\partial t}+\dfrac{1}{c}\br J\label{eq:im2}
\end{equation}
\begin{equation}
\bg\nabla\cdot\br B=0\:\:\quad-\bg\nabla\times\br E=\dfrac{1}{c}\dfrac{\partial\br B}{\partial t}\label{eq:im3}
\end{equation}
Now apply the dyality transformation in \prettyref{eq:tx1}. The extended
Maxwell equations then become
\begin{equation}
-\bg\nabla\cdot\br B'=-\lambda'\quad\:\:\bg\nabla\times\br E'=-\dfrac{1}{c}\dfrac{\partial\br B'}{\partial t}-\dfrac{1}{c}\br L'\label{eq:im4}
\end{equation}
\begin{equation}
\bg\nabla\cdot\br E'=0\:\:\quad\bg\nabla\times\br B'=\dfrac{1}{c}\dfrac{\partial\br E'}{\partial t}\label{eq:im5}
\end{equation}
 in which $\fv L'\neq0$. Even if we were to begin with a solution
with $\fv L=0$, corresponding to the current experimental evidence
that no magnetic charge exists, an active transformation such as \prettyref{eq:tx1}
would imply the real existence of an alternate solution in which $\fv L'\neq0$.
With the active interpretation of the dyality transformation, eqs.(\ref{eq:im4},
\ref{eq:im5}) represent an experimentally real alternate solution
in which magnetic charge is non-zero. Thus the active interpretation
is ruled out currently by the failure to find those solutions experimentally.

However, a \emph{passive }interpretation of dyalitic transformations
can be used. In the \emph{passive interpretation,} the primed quantities
in \prettyref{eq:tx1} are interpreted as representing the same physical
reality as the unprimed ones, just viewed differently. 

The analogy here is with passive rotations in standard three-dimensional
vector algebra. The coordinate system is rotated in the opposite sense,
and the vector components transform as before, by $V'_{i}=\sum R_{ij}V_{j}$.
But the vector $\br V$ itself is unchanged. Vector $\br V$ has different
components only because it is being viewed now from a different observer
orientation.

For example, the passive interpretation holds \prettyref{eq:im1a}
and \prettyref{eq:im1b} to represent the same physical reality. The
only change is that quantities previously denoted $\br E$, $\br B$
are now being denoted by $-\br B'$, $\br E'$. In the other example,
with the passive interpretation of this dyality transformation, eqs.(\ref{eq:im4},
\ref{eq:im5}) represent the same physical reality as eqs.(\ref{eq:im2},
\ref{eq:im3}). The only change is that quantities previously denoted
$\br E$, $\br B$, $\fv J$ are now being denoted by $-\br B'$,
$\br E'$, $-\fv L'$. 

The passive interpretation is evidently used by Jackson\footnote{Section 6.12, p.252 of \citep{Jackson}. }
when he says that, if all\footnote{Note the word \emph{all}; universality is required. Clearly, such
a scheme will only be coherent if it is universal. All of physics
would have to agree on the chosen $\chi$ value. } particles have the same ratio of magnetic to electric charge, then
we are free to use the generalized definition of dyalitic transformation
in Appendix B and choose $\chi$ in \prettyref{eq:b1} such that $\fv L'=0$.
Call that ratio $r=q_{m}/q_{e}$, and use the second of \prettyref{eq:b1}
to write
\begin{equation}
\fv L'=\fv J\left(-\sin\chi+r\cos\chi\right)\label{eq:im6}
\end{equation}
and then choose $\tan\chi=r$ to make $\fv L'=0$. Then we \textquotedbl{}...
have the Maxwell equations as they are usually known.\textquotedbl{}
In that case, \textquotedbl{}...it is a matter of convention to speak
of a particle possessing an electric charge, but not magnetic charge.\textquotedbl{}

And, at present there is indeed a universally accepted value for the
ratio $r$ of magnetic to electric charge of all particles. It is
$r=0$. Thus, viewing the situation using Jackson's language, we are
free to begin with $\fv L=0$ and use $r=0$ and $\chi=0$ (the identity
transformation) to maintain $\fv L'=0$ uniformly and consistently.
Just as we choose our coordinate system for convenience in standard
vector calculus, we are free to choose $\chi=0$ for our convenience.
The result is still extended electrodynamics, but with its passive
dyality symmetry used to justify maintaining $\fv L'$ equal to zero.
Thus the standard form of Maxwell equations in eqs.(\ref{eq:im2},
\ref{eq:im3}) can be considered as extended Maxwell equations with
passive dyality symmetry used to maintain $\chi=0$ and $\fv L=0$
universally.

However, if a particle with a magnetic charge ratio $\left|q_{m}/q_{e}\right|$
greater than the near-zero value of the electron is eventually found,
then there will no longer be a \emph{universal }value of $r$. It
will have a near-zero value for the electron, and a presumably larger
value for the magnetically charged particle. Then passive dyality
invariance can no longer be used to keep $\fv L'=0$ universally.

Thus, if some future magnetic-charge-detection experiment can be described
correctly by the extended Maxwell equations with zero magnetic charge
for the electron but nonzero magnetic charge for some other particle,
then there will no longer be a passive dyality transformation that
transforms away the magnetic charge sources. If nonzero magnetic charge
exists, experimental evidence for its existence cannot be hidden by
a passive dyality transformation.

\section{Afterword\label{sec:Afterword}}

It follows from \prettyref{eq:em22} that the Poynting vector $\br S=c\br E\times\br B$
is the flux density of the field energy ${\cal E}=\frac{1}{2}\left(E^{2}+B^{2}\right)$.
If one were to suppose that electromagnetic energy flows as a classical
fluid, then there would be a velocity $\br V(x)$ defined at each
field point such that\footnote{See Chapter 2 of \citep{batchelor}.}
\begin{equation}
\br S={\cal E}\br V\label{eq:aft1}
\end{equation}
If that were the case, then dividing by $c^{2}$ and using the definition
of field momentum $\br P$ from \prettyref{eq:intro4} would give
\begin{equation}
\br P={\cal M}\br V\label{eq:aft2}
\end{equation}
 where ${\cal M}={\cal E}/c^{2}$ is a relativistic mass density.
We could then conclude that field momentum is due to the flow of field
mass.

However, it has been proved\footnote{See \citep{johns-energyflow}.}
that there is no special relativistically correct velocity $\br V$
satisfying \prettyref{eq:aft1}. This is because the conservation
of electromagnetic energy-momentum is expressed as the divergence
of a symmetric four-tensor $T^{\alpha\beta}$ rather than as the divergence
of a (non-existent) four-vector. 

Thus \prettyref{eq:aft2} is not true; electromagnetic field momentum
cannot be explained as due to the flow of relativistic mass. And yet
electromagnetic field momentum is freely exchangeable with particle
momentum. If electromagnetic field momentum is not due to the flow
of field energy, then what is it? And what is the particle momentum
with which it can be freely exchanged?

Attempts to answer those questions may provide clues to a future quantum
mechanics.\footnote{For example, the non-existence of $\br V(x)$ in \prettyref{eq:aft2}
may relate to the non-existence of velocity eigenstates in quantum
mechanics.} If so, we should concentrate on a deeper understanding of electrodynamics,
as this paper attempts to encourage.

Also, the deficiencies of the formal Lagrangian derivation of extended
electrodynamics noted in \prettyref{sec:Lagrange} suggest that future
fundamental electrodynamic research should concentrate on the extended
Maxwell equations themselves and not on their Lagrangian derivation.

\section{Appendix A\label{sec:AppendixA}:The Poynting Theorem}

To prove that $\partial_{\alpha}T^{\alpha\beta}=-f^{\beta}=-\left(f_{\text{el}}^{\beta}+f_{\text{mg}}^{\beta}\right)$.

From \prettyref{eq:du13}, this is to prove that
\begin{equation}
-\left(f_{\text{el}}^{\beta}+f_{\text{mg}}^{\beta}\right)=\partial_{\alpha}T^{\alpha\beta}=\frac{1}{2}\partial_{\alpha}\left\{ F^{\alpha\mu}F^{\beta\nu}\,\eta_{\mu\nu}+G^{\alpha\mu}G^{\beta\nu}\,\eta_{\mu\nu}\right\} \label{eq:ap1}
\end{equation}
which, using \prettyref{eq:em19} and \prettyref{eq:em20} and the
product rule, is equivalent to
\begin{equation}
\left\{ F_{\:\:\gamma}^{\beta}\left(\partial_{\alpha}F^{\alpha\gamma}\right)+G_{\:\:\gamma}^{\beta}\left(\partial_{\alpha}G^{\alpha\gamma}\right)\right\} =\dfrac{1}{2}\Big\{ F_{\:\:\gamma}^{\beta}\left(\partial_{\alpha}F^{\alpha\gamma}\right)+F_{\:\:\gamma}^{\alpha}\left(\partial_{\alpha}F^{\beta\gamma}\right)+G_{\:\:\gamma}^{\beta}\left(\partial_{\alpha}G^{\alpha\gamma}\right)+G_{\:\:\gamma}^{\alpha}\left(\partial_{\alpha}G^{\beta\gamma}\right)\Big\}\label{eq:ap1a}
\end{equation}

Thus, we must prove that
\begin{equation}
F_{\:\:\gamma}^{\beta}\left(\partial_{\alpha}F^{\alpha\gamma}\right)+G_{\:\:\gamma}^{\beta}\left(\partial_{\alpha}G^{\alpha\gamma}\right)=F_{\:\:\gamma}^{\alpha}\left(\partial_{\alpha}F^{\beta\gamma}\right)+G_{\:\:\gamma}^{\alpha}\left(\partial_{\alpha}G^{\beta\gamma}\right)\label{eq:ap2}
\end{equation}
 Defining
\begin{equation}
\phi_{\beta}\equiv G_{\beta\gamma}\left(\partial_{\alpha}G^{\alpha\gamma}\right)-G^{\alpha\gamma}\left(\partial_{\alpha}G_{\beta\gamma}\right)\label{eq:ap3}
\end{equation}
we must prove that
\begin{equation}
\phi_{\beta}=-F_{\!\!\beta\gamma}\left(\partial_{\alpha}F^{\alpha\gamma}\right)+F^{\alpha\gamma}\left(\partial_{\alpha}F_{\!\!\beta\gamma}\right)\label{eq:ap4}
\end{equation}
 Inserting \prettyref{eq:du1} into \prettyref{eq:ap3} gives
\[
\phi_{\beta}=\frac{1}{4}\varepsilon_{\gamma\beta\theta\delta}F^{\theta\delta}\left(\partial_{\alpha}\varepsilon^{\gamma\alpha\mu\nu}F_{\!\!\mu\nu}\right)-\frac{1}{4}\varepsilon^{\gamma\alpha\mu\nu}F_{\!\!\mu\nu}\left(\partial_{\alpha}\varepsilon_{\gamma\beta\theta\delta}F^{\theta\delta}\right)
\]
 Using the identity \prettyref{eq:use4}, this may be written
\begin{equation}
\phi_{\beta}=\frac{1}{4}\Big\{-\left(\delta_{\beta}^{\alpha}\delta_{\theta}^{\mu}\delta_{\delta}^{\nu}+\delta_{\theta}^{\alpha}\delta_{\delta}^{\mu}\delta_{\beta}^{\nu}+\delta_{\delta}^{\alpha}\delta_{\beta}^{\mu}\delta_{\theta}^{\nu}\right)+\left(\delta_{\beta}^{\alpha}\delta_{\delta}^{\mu}\delta_{\theta}^{\nu}+\delta_{\delta}^{\alpha}\delta_{\theta}^{\mu}\delta_{\beta}^{\nu}+\delta_{\theta}^{\alpha}\delta_{\beta}^{\mu}\delta_{\delta}^{\nu}\right)\Big\}\left\{ F^{\theta\delta}\partial_{\alpha}F_{\!\!\mu\nu}-F_{\!\!\mu\nu}\partial_{\alpha}F^{\theta\delta}\right\} \label{eq:ap6}
\end{equation}
\begin{equation}
\phi_{\beta}=-\frac{1}{2}\left(\delta_{\beta}^{\alpha}\delta_{\theta}^{\mu}\delta_{\delta}^{\nu}+\delta_{\theta}^{\alpha}\delta_{\delta}^{\mu}\delta_{\beta}^{\nu}+\delta_{\delta}^{\alpha}\delta_{\beta}^{\mu}\delta_{\theta}^{\nu}\right)\left(F^{\theta\delta}\partial_{\alpha}F_{\!\!\mu\nu}-F_{\!\!\mu\nu}\partial_{\alpha}F^{\theta\delta}\right)\label{eq:ap7}
\end{equation}
\begin{equation}
\phi_{\beta}=-\frac{1}{2}\left(F^{\mu\nu}\partial_{\beta}F_{\!\!\mu\nu}+F^{\alpha\mu}\partial_{\alpha}F_{\!\!\mu\beta}+F^{\nu\alpha}\partial_{\alpha}F_{\!\!\beta\nu}\right)+\frac{1}{2}\left(F_{\!\!\mu\nu}\partial_{\beta}F^{\mu\nu}+F_{\!\!\mu\beta}\partial_{\alpha}F^{\alpha\mu}+F_{\!\!\beta\nu}\partial_{\alpha}F^{\nu\alpha}\right)\label{eq:ap8}
\end{equation}
\begin{equation}
\phi_{\beta}=-F_{\!\!\beta\gamma}\left(\partial_{\alpha}F^{\alpha\gamma}\right)+F^{\alpha\gamma}\left(\partial_{\alpha}F_{\!\!\beta\gamma}\right)\label{eq:ap9}
\end{equation}
which is \prettyref{eq:ap4}, as was to be proved.

\section{Appendix B\label{sec:AppendixB}: Generalized Dyality Transformation}

The definition of dyality transformation in \prettyref{eq:tx1} can
be generalized to
\begin{align}
\left(\begin{array}{c}
\br E'\\
\br B'
\end{array}\right) & =\left(\begin{array}{cc}
\cos\chi & \sin\chi\\
-\sin\chi & \cos\chi
\end{array}\right)\left(\begin{array}{c}
\br E\\
\br B
\end{array}\right)\nonumber \\
\left(\begin{array}{c}
\fv J'\\
\fv L'
\end{array}\right) & =\left(\begin{array}{cc}
\cos\chi & \sin\chi\\
-\sin\chi & \cos\chi
\end{array}\right)\left(\begin{array}{c}
\fv J\\
\fv L
\end{array}\right)\label{eq:b1}\\
\left(\begin{array}{c}
\fv A'\\
\fv N'
\end{array}\right) & =\left(\begin{array}{cc}
\cos\chi & \sin\chi\\
-\sin\chi & \cos\chi
\end{array}\right)\left(\begin{array}{c}
\fv A\\
\fv N
\end{array}\right)\nonumber 
\end{align}
where $\chi$ is a constant parameter.\footnote{See eqs.(6.151 and 6.152) of \citep{Jackson}.}
The earlier definition in \prettyref{eq:tx1} is the special case
with $\chi=\pi/2$. Due to the linearity and orthogonality of \prettyref{eq:b1},
all dyality invariance arguments given earlier in the paper using
$\chi=\pi/2$ are also valid for general $\chi$. For example, \prettyref{eq:tx6}
generalizes to
\begin{equation}
\left(\begin{array}{c}
F'^{\alpha\beta}\\
G'^{\alpha\beta}
\end{array}\right)=\left(\begin{array}{cc}
\cos\chi & \sin\chi\\
-\sin\chi & \cos\chi
\end{array}\right)\left(\begin{array}{c}
F^{\alpha\beta}\\
G^{\alpha\beta}
\end{array}\right)\label{eq:b2}
\end{equation}
Also $T'^{\alpha\beta}=T^{\alpha\beta}$ and the Poynting theorem
in \prettyref{eq:em21} is preserved, regardless of the $\chi$ value.

\end{document}